\documentclass[fleqn,usenatbib]{mnras}
\usepackage[T1]{fontenc}
\usepackage{ae,aecompl}
\usepackage{graphicx}   
\usepackage{amsmath}    
\usepackage{amssymb}    

\usepackage[dvipsnames]{xcolor}

\title[Fast Prograde Flows]{Fast Prograde Coronal Flows in Solar Active Regions}

\author[H.S. Hudson et al.]{
\begin{tabular}{lll}
Hugh S. Hudson,$^{1,2}$\thanks{hugh.hudson@glasgow.ac.uk}
Sargam M. Mulay,$^{1}$
Lyndsay Fletcher,$^{1,3}$\\
Jennifer Docherty,$^{1}$
Jimmy Fitzpatrick,$^{1}$
Eleanor Pike,$^{1}$
Morven Strong,$^{1}$\\
Phillip C. Chamberlin,$^{4}$, and 
Thomas N. Woods$^{4}$
\end{tabular}
\bigskip
\\
$^{1}$SUPA School of Physics and Astronomy, University of Glasgow, Glasgow G12 8QQ UK\\
$^{2}$Space Sciences Laboratory, University of California, Berkeley, CA 94720 USA\\
$^{3}$Rosseland Centre for Solar Physics, University of Oslo, P.O.Box 1029, Blindern, 03015 Oslo, Norway\\
$^{4}$LASP, Boulder CO 80303 USA
}

\date{Accepted XXX. Received YYY; in original form ZZZ}
\pubyear{WWW}

\begin{document}

\label{firstpage}
\pagerange{\pageref{firstpage}--\pageref{lastpage}}
\maketitle

\maketitle

\begin{abstract}
We report the discovery and characterization of high-speed ($>100$~km/s) horizontal flows in solar active regions, making use of the Sun-as-a-star spectroscopy in the range 5-105~nm provided by the EVE(Extreme Ultraviolet Variability Experiment) spectrometers on the \textit{Solar Dynamics Observatory}.
These apparent flows are persistent on time scales of days, and are well observed in lines of Mg~{\sc x}, Si~{\sc xii} and Fe~{\sc xvi} for example.
They are prograde, as evidenced directly by blueshifts/redshifts peaking at the east/west limb passages of isolated active regions.
The high-speed flow behavior does not depend upon active-region latitude or solar cycle, with similar behavior in Cycles~24 and~25.
\end{abstract} 


\section{Introduction}\label{sec:intro}

Spectroscopy of the solar corona in the extreme ultraviolet (EUV) can detect coronal flows via the Doppler effect, and considerable literature exists based on the CDS\footnote{Coronal Diagnostic Spectrometer (SOHO)}, SUMER\footnote{Solar Ultraviolet Measurements of Emitted Radiation (SOHO)}, and EIS\footnote{EUV imaging spectrometer (\textit{Hinode})} instruments in particular.
For reviews see for example \citet{2014LRSP...11....4R} and \cite{2018LRSP...15....5D}.
Particular attention has focused on vertical (radial) motions that might relate to the formation of the solar wind (``open'' fields) and loop-like structures (``closed'' fields), and with transient flows associated with flares.
These high-resolution instruments use scanning slits, with limited fields of view and incomplete time-domain coverage.
However, mosaic tiling can produce global Doppler maps such as those shown by \cite{2015NatCo...6.5947B}; their example took two days of observation to accumulate and, on a global scale, generally suggested patchy line-of-sight flows below $\pm 15$~km/s in Fe~{\sc xiii} 20.2~nm, nominally formed at log~\textit{T} [K]~=~6.25.

This paper reports on global flow properties, as distinguished from transient effects. 
We find that quiet-Sun Doppler shifts at active-region temperatures show a persistent prograde zonal character that had not been recognized by the earlier imaging-spectroscopy instruments.
Our observations come from the SDO/EVE/MEGS\footnote{Extreme Ultraviolet Experiment/Multiple Grating Spectograph} instrument which produces Sun-as-a-star spectra of many emission lines formed throughout the solar atmosphere.
We specifically use the MEGS-B spectrometer \citep{2012SoPh..275..115W}, which has coverage mostly of  of the chromosphere and transition region, but also a few coronal lines over a spectral band approximately 33-105~nm.
These data have remarkable stability, continuity, and sensitivity; \cite{2011SoPh..273...69H} showed that the He~{\sc ii} 30.4 nm line readily detected the SDO orbital motion of 6.1~km/s peak-to-peak, with high significance. 
EVE could also detect flare-related Doppler shifts, even at its basic 10-s sampling.
Note that the orbital diurnal Doppler amplitude is only 0.6~pm as compared with the 40~pm spectral binning: the Doppler observations studied here correspond to subtle shifts of the line centroids well below the scale of the bin width or of the instrumental line width.
It was clear nevertheless from these early characterizations \citep[see also][]{2016A&A...596A..51B,2016SoPh..291.1665C} that the Doppler measurements were robust, and that variability on longer time scales (i.e., the non-flaring Sun) was also present.
Recent observations of the flare SOL2021-10-28 also found high-speed ejecta (of order 500~km/s)  in MEGS-B Doppler observations of transition-region lines \citep{2022ApJS..260...36Y}.

This paper describes the first exploration of longer time scales ($\geq 1$~h) in Sun-as-a-star Doppler observations not associated with flares.
The results are global in nature, because of EVE's Sun-as-a-star character, but we can identify the observed flows with active regions in detail because of the occurrence of isolated regions.
It is worth noting that many earlier observations have detected flows in EUV lines, but not what EVE shows us.
\cite{1987ApJ...323..368K} summarizes some of the earlier work with ``...no clear and complete picture of active region dynamics has yet emerged...''
In view of our new results, this may still be a reasonable assessment.
The EIS instrument \cite{2007SoPh..243...19C} on \textit{Hinode} in particular should have the capability to detect what we report below.
We describe the resulting tension below in Section~\ref{sec:tension}.

\section{Observations}\label{sec:obs}

\subsection{Overview}

The data we discuss in this paper come entirely from the MEGS-B spectrograph of the EVE instrument on board SDO. 
The MEGS-B design has two fixed concave gratings with a CCD camera at the focal plane, and potential wavelength shifts from thermal gradients are reduced by tight control of the MEGS-B temperature range to less than 1~C.
We do not use MEGS-A because of astigmatism in the optics, which produces wavelength shifts comparable to the ones we study; at present these artifacts cannot be corrected \citep{2007SPIE.6689E..18C,2016SoPh..291.1665C}.

MEGS-B has no imaging capability and observes Sun-as-star spectra with a resolution of 1~\AA\ and a cadence of 10~s initially (from April 2010), reduced to 60~s in 2019 to improve the signal-to-noise ratio of the data.
The geostationary orbit of SDO gives these data continuity on any time scale in principle, but in practice the slow degradation of the MEGS-B optics requires that it be operated only intermittently, with varying patterns over the years of observation, and almost always with some data each day.
This does not affect the results presented in this paper, because daily sampling suffices to characterize the global Doppler patterns we have found.

Unwanted diurnal noise provides another systematic issue for the EVE/MEGS data.
This arises in the Level-2 and Level-2B data products from a correction made to minimize the effects of background electrons, which can be detected at CCD locations in the instrument not used for spectra.
The electron fluxes have a diurnal pattern, maximizing in the antisolar direction when SDO is slightly penetrating the radiation belts or geotail.
The correction has a subtle effect on the EUV wavelength calibrations, and different lines show different patterns.
Early in the mission this noise term was not large, but it grew with time and has not been studied extensively.
Mitigation would be outside the scope of the work here; we avoid most of the issue by simply sampling the data in the anti-geotail direction at local SDO perihelion times (17:00-18:00~UT, corresponding to local noon for the geosynchronous downlink station in New Mexico).

Generally our data reduction consists of making standard single-Gaussian fits, including a quadratic background level for each line (6 free parameters).
At the 1~\AA\ resolution of EVE, blends frequently occur but are minimally important for the 
MEGS-B line list shown in Table~\ref{tab:linelist}. 
The results in this paper further avoid confusing blends by using multiple redundant lines for confirmation.

\begin{table}
\centering
\caption{MEGS-B non-flare line list}
\label{tab:linelist}
\begin{tabular}{r r l r}
\hline
Wavelength, \AA\ & Species & Ion & log~\textit{T} [K]\\
\hline
  335.41 &   Fe &    XVI &     6.49 \\
  360.76 &   Fe &    XVI &     6.49 \\
  368.07 &   Mg &     IX &     6.00 \\
  434.92 &   Mg &    VII &     5.80 \\
  436.73 &   Mg &   VIII &     5.90 \\
  439.18 &   Mg &     IX &     6.05 \\
  445.70 &    S &    XIV &     6.50 \\
  465.22 &   Ne &    VII &     5.70 \\
  499.41 &   Si &    XII &     6.30 \\
  521.00 &   Si &    XII &     6.30 \\
  525.79 &    O &    III &     4.90 \\
  537.03 &   He &      I &     5.52 \\
  554.51 &    O &     IV &     5.20 \\
   584.34 &   He &      I &     5.44 \\
   599.59 &    O &    III &     5.00 \\
  609.80 &   Mg &      X &     6.07 \\
  624.94 &   Mg &      X &     6.07 \\
  629.73 &    O &      V &     5.35 \\
  718.53 &    O &     II &     4.45 \\
   770.41 &   Ne &   VIII &     5.81 \\
  790.20 &    O &     IV &     5.20 \\
  835.50 &    O &     II &     4.52 \\
  949.70 &    H &      I &     4.0 \\
  972.54 &    H &      I &     4.0 \\
  977.16 &    C &    III &     4.85 \\
  991.51 &    N &    III &     4.95 \\
 1025.72 &    H &      I &     4.0 \\
 1031.91 &    O &     VI &     5.5 \\
 \hline
\end{tabular}
\end{table}

Because of the novelty of these data, and the tension they present relative to the literature (Section~\ref{sec:tension}) we briefly discuss possible systematic errors in our database here.
First, EVE has relatively low spectral resolution and at the level of precision used here, both known and unknown blends might confuse the Doppler measurements.
For a given line, an unwanted blend would indeed produce a spurious Doppler shift, but it would be the same sense at either limb, and thus cannot contribute to the asymmetry we observe.
Second, unstable wavelength calibrations might conceivably occur, but cool lines not showing the flows demonstrate great stability. 
Lines showing large Doppler shifts (hot) and lines not showing them (cool) occur near each other in wavelength.
We do not consider the effects of finite optical depth, but the flows do occur in lines of varying strengths.
Finally, existing EUV data do show blue asymmetries and upflows in specific sites, but with net velocities too small to explain the EVE observations.

\subsection{The 2010 epoch}

\begin{figure}
\centering
   \includegraphics[width=0.49\textwidth]{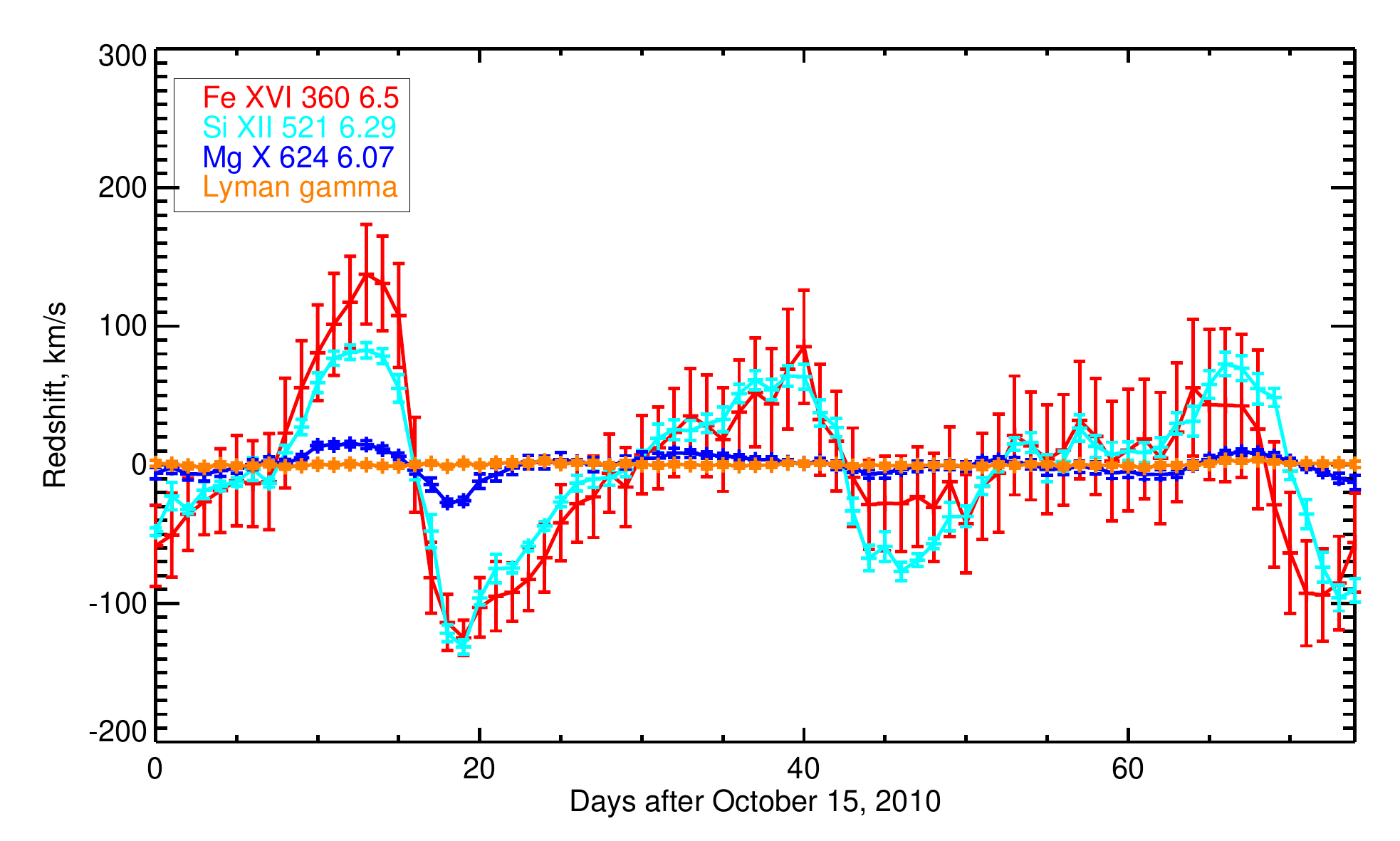}
           \caption{\textit{Relative Doppler shifts for several MEGS-B lines, taking zero velocity as the timeseries median in each case. The error bars here are from the Gaussian line-fitting and are larger than the statistical uncertainties in all cases. Each point represents an integration from 17:00-18:00~UT to avoid minor diurnal systematic errors (see text). The legend names the lines and (except for Lyman-$\gamma$) gives their wavelengths in \AA, as well as the log~(\textit{T}). 
           The figure legend lists the lines in the order they appear in the major excursion at about Day~13.        
           }
       }
\label{fig:show_doppler}
\end{figure}

Figure~\ref{fig:show_doppler} shows an example of the observations over a 11-week time span at the end of 2010, for wavelengths listed in the Legend with wavelengths (\AA) and  log~\textit{T} estimates.
This time interval immediately followed the Cycle 23/24 solar minimum and was during the first year of SDO operations.
It further included isolated active regions from which we could infer Doppler shifts without ambiguities despite EVE's lack of imaging.
The Figure reveals a clear pattern of Doppler variation on time scales characteristic of active-region development and of solar rotation.
The error bars show only the parameter uncertainties for the Gaussian fit for each line profile, and thereby significantly exceed the statistical errors because the instrumental line profiles are only approximately Gaussians.

The hotter lines have larger amplitudes in these large-scale Doppler variations, but the absolute velocities remain to be determined.
The reason for this uncertainty is the lack of absolute wavelength calibrations at the accuracy needed.
This uncertainty does not affect our conclusions in any way because we report only differential Doppler shifts; the stability of these relative measurements is guaranteed by deep atmospheric lines (such as Lyman-$\gamma$) that do not show the high-speed flows.
Re-calibration to avoid these perturbations is beyond the scope of this paper. 
The actual vector velocities in any case can only be determined up to the limit imposed by geometrical projection effects (see further discussion below in Section~\ref{sec:interp}).

\begin{figure}
\centering
   \includegraphics[width=0.49\textwidth]{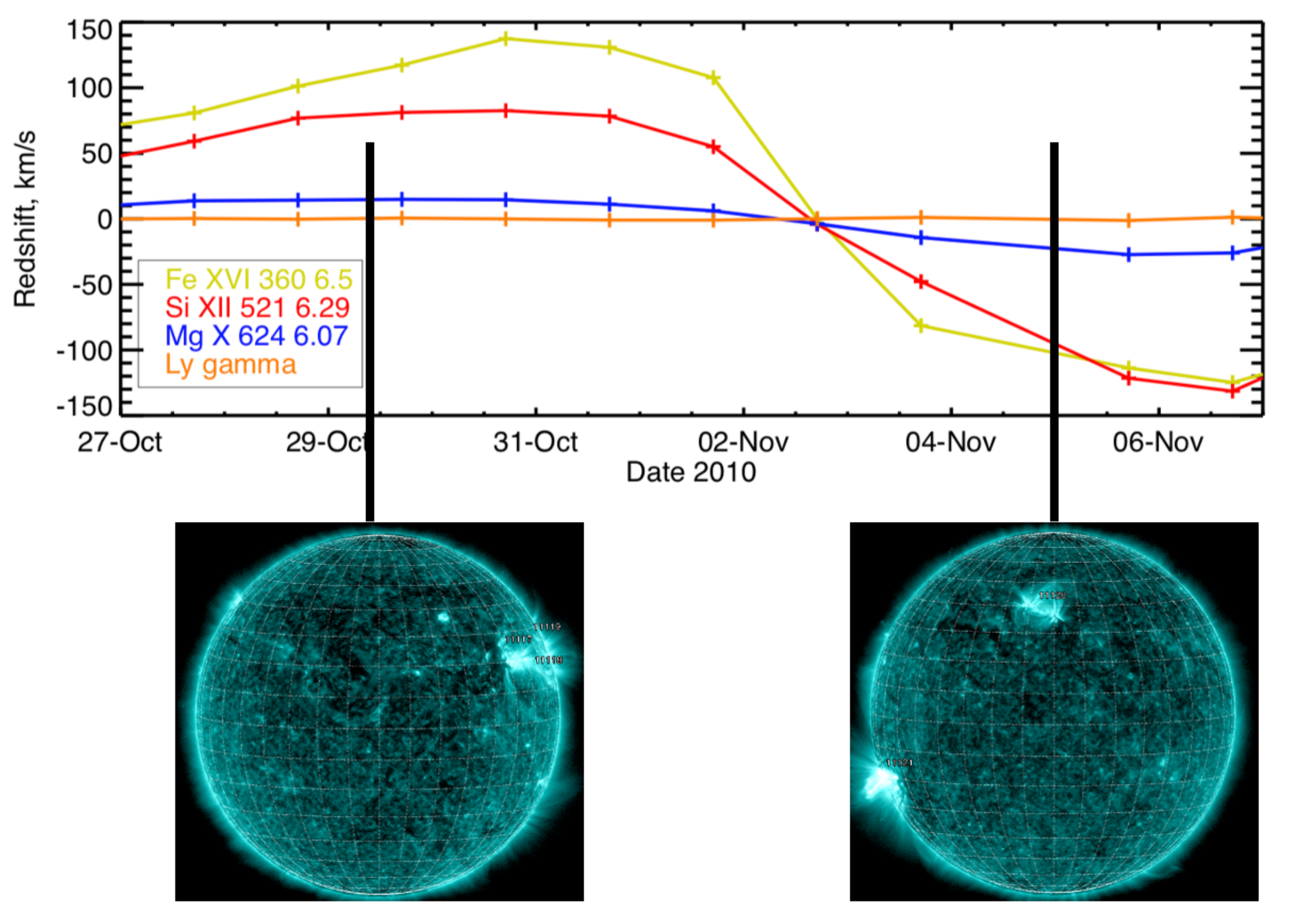}
           \caption{\textit{The early part of the timeseries of Figure~\ref{fig:show_doppler}, with sample frames from the AIA 94~\AA\ channel to illustrate the active-region populations resulting in the Doppler excursions.}
       }
\label{fig:regions_graphic}
\end{figure}

The association of these global Doppler shifts with active regions is well established by Figure~\ref{fig:show_doppler}; the decrease of the large relative redshift in early November 2010 coincides with gradual disappearance of NOAA region 11117 (central meridian passage 25 October, spot area about 500~MSH\footnote{Millionths of the Solar Hemisphere} at occultation).
This was immediately followed by the SE limb appearance of regions 11121, 11124, and 11125, corresponding to the observed relative blueshift.
We emphasize this behavior in the graphic of Figure~\ref{fig:regions_graphic}.
Other isolated regions, not discussed here, establish that the flow direction does not depend upon latitude.

\begin{figure}
\centering
   \includegraphics[width=0.49\textwidth]{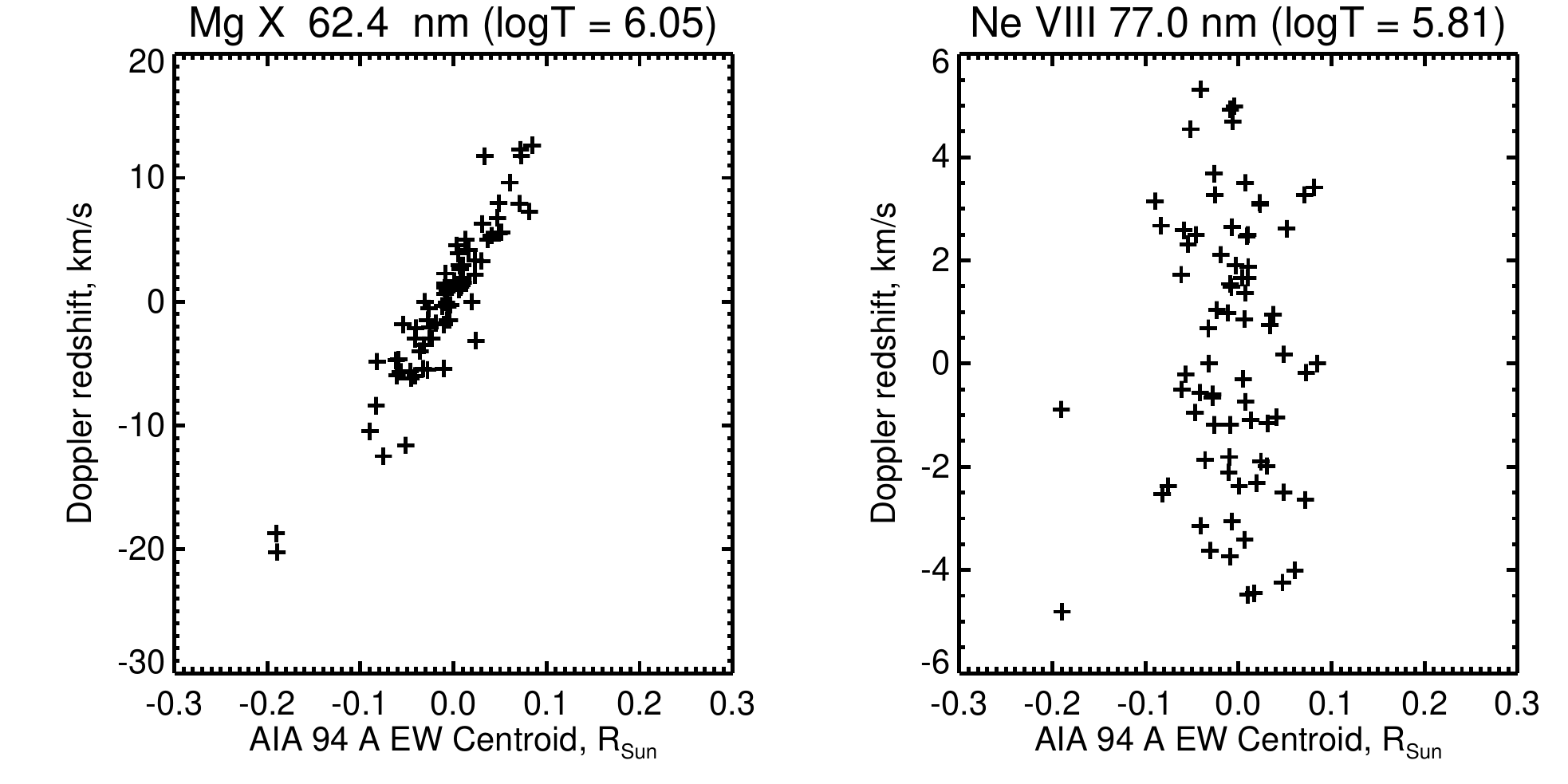}
           \caption{\textit{Correlations between EVE Doppler residuals and AIA image X-centroids.
           Left, Mg~{\sc x} 62.5~nm; right, Ne~{\sc viii} 770~nm .}
       }
\label{fig:flow_grade_2panel}
\end{figure}

For the same time range, we have also obtained daily AIA 94~\AA\ images and extracted the east-west (EW) image centroids by calculating the intensity-weighted mean
$$\bar{X} = \frac{\sum{X \times I(x,y)}}{\sum{I(x,y)}}\ \ ,$$
where $I(x,y)$ is the image and $X$ is the helioprojective EW angle normalized to $R_\odot$.
This leads to the correlation between EVE Doppler shift of Mg~{\sc x} (log(\textit{T})~=~6.05) and AIA image structure seen in the right panel of Figure~\ref{fig:flow_grade_2panel}.
Interestingly the Ne~{\sc viii} line (77~nm; log(\textit{T})~=~5.8) and other transition-region lines do not show this correlation.
At present our knowledge of the temperature dependence of the flow speeds is limited, but the Figure shows a striking distinction between the behavior of  a coronal line at the core temperature of an active region, and that of a transition-region line still at a relatively high formation temperature.

\subsection{Later times}

\begin{figure}
\centering
   \includegraphics[width=0.49\textwidth]{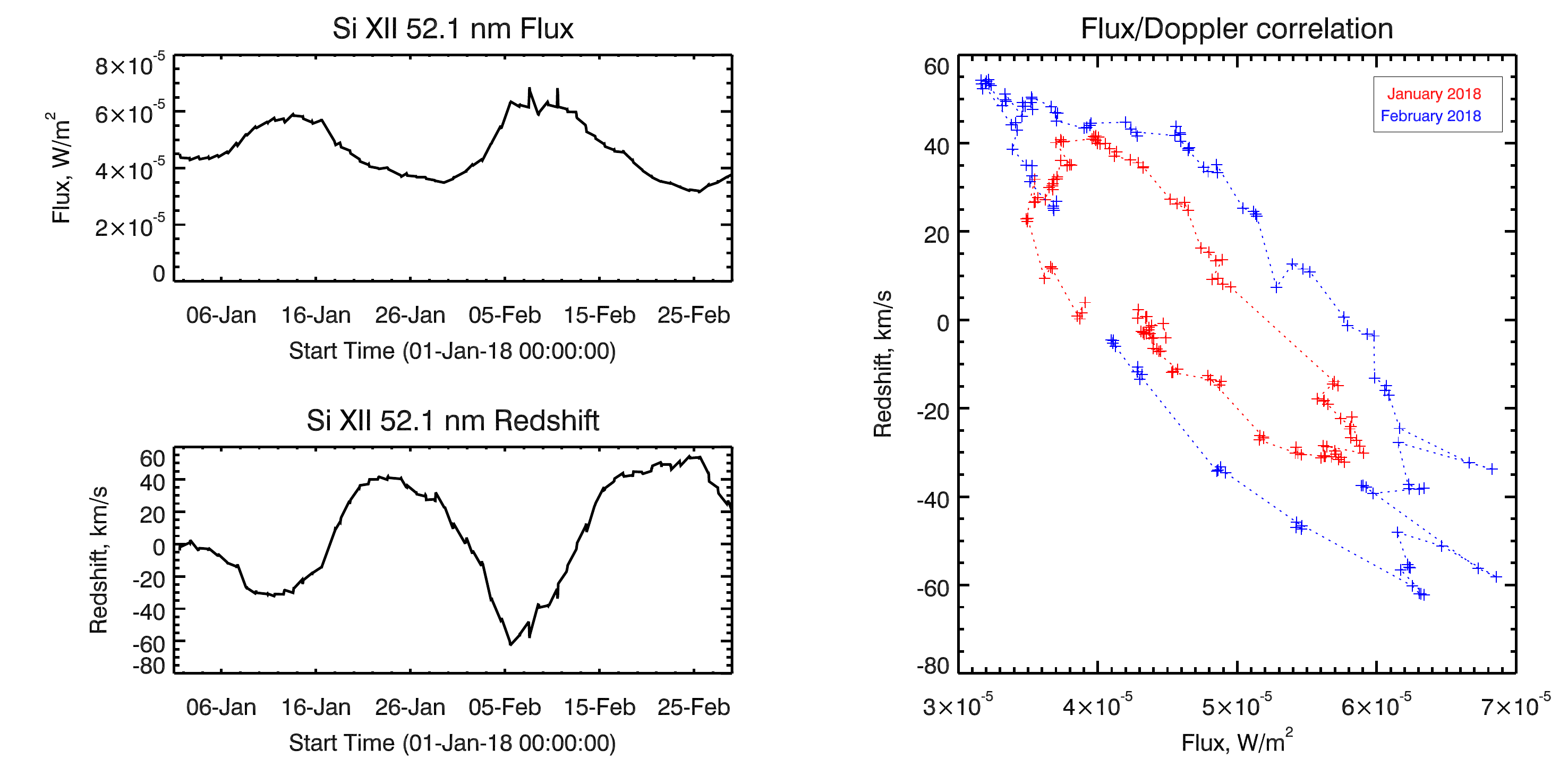}
           \caption{\textit{The Doppler/image correlation in a different format, and now for two solar rotations of data from 2018. Left, the Si~{\sc xii} 52.1~nm flux and Doppler timeseries; right, their clockwise correlations for the two rotations (red inner loop from January). Again, the phase shift is that expected for prograde flows.}
       }
\label{fig:correlations_2018}
\end{figure}

We have searched extensively over the EVE database, continuing now into Cycle 25.
The properties found in the 2010 data described above continue, independent of region latitude or cycle number.
Although we have not made complete detailed studies, it appears that the measured component of the flow has a net prograde sense (along lines of heliolatitude in the direction of solar rotation). 
We have not yet searched for a dependence on the tilt angle, but this would be informative.
Figure~\ref{fig:correlations_2018} shows another example of the Doppler/image correlation for a period in 2018, using a slightly different format.
Here we illustrate the phase lag between AR brightness (disk passage) and Doppler signature (limb passage), finding as expected a roughly 90$^\circ$ phase lag for redshifts.

\section{Interpretation}\label{sec:interp}

The observed Doppler shifts are lower limits for the horizontal flows, which we observe in projection.
A further limiting  factor is that AIA images and monochromatic EIS images \citep[e.g.,][]{2015NatCo...6.5947B} show large contributions from the general corona and multiple active regions, which will dilute the EVE line shifts.
Both projection and dilution effects thus tend to make our Doppler speeds lower limits.

The prograde sense of the observed flows matches the expectation for siphon action between stronger leading-polarity field regions and weaker following fields.
However existing models \citep{1980SoPh...65..251C,1995A&A...294..861O} do not suggest any relevance to the observed high-speed flows.

We suspect that most closed active-region magnetic fields support the pattern that EVE detects, because of its pervasive presence of the Doppler signatures in our Sun-as-a-star data. 
This has implications for structural dynamics and for the generation of turbulence, as well as for general scaling laws \citep{1995A&A...294..861O}.
As noted by \cite{1999ApJ...517L.155L}, we would not even detect these loops if they did not contain hot plasma.
Future theory and modeling of the hot core loops of active regions will need to deal with the relationship between temperature and flow speed.

\section{Can we reconcile EIS and EVE Doppler observations?}\label{sec:tension}

\begin{figure}
\centering           
\includegraphics[width=0.49\textwidth]{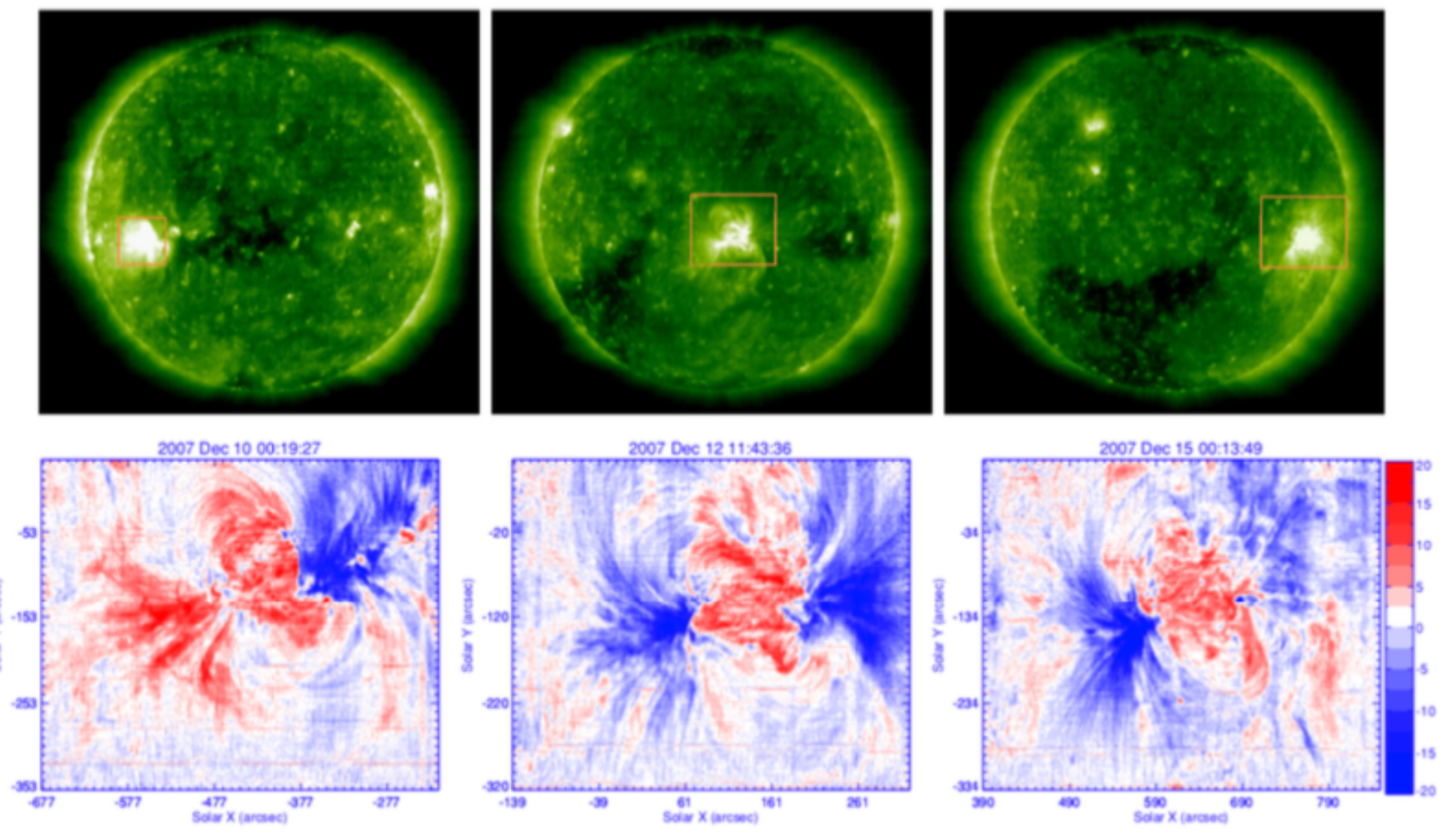}
\caption{\textit{A glimpse of the Doppler coverage of AR~10798 (December 2007) as observed by 
EIS in Fe~{\sc xii} 195~\AA\ (logT = 6.1), adapted from Bryans \textit{et al.} 2010.
The Doppler color table (lower panels) has a range $\pm 15$~km/s.
Without reference to the details, this is sufficent to highlight the tension between EIS and EVE (see text)
}
        }
\label{fig:bryans}
\end{figure}

Why have the scanning-slit EUV spectrometers not detected  the flow patterns described above?
In particular the \textit{Hinode}/EIS instrument immediately had an excellent opportunity to do so, with its early observations of the nicely isolated major active region NOAA 10978  tracked across its disk passage in December 2007 and the subject of many studies \citep{2008ApJ...678L..67H,2010ApJ...715.1012B,2010A&A...521A..51P,2012ApJ...760L...5B}.
For the most part these studies have emphasized high-speed blueshifts as possible contributors to the solar wind along open fields.
Figure~\ref{fig:bryans} illustrates the Bryans \textit{et al.} data for Fe~{\sc xii} 195~\AA\ (log~T~=~6.1), and at a glance one cannot see the systematic high-speed flows expected from our EVE data; we would expect to see predominant blueshifts at the east and predominant redshifts at the west; note also that the entire scale of the Doppler shifts in this Figure is only $\pm 15$~km/s.

\begin{figure}
\centering           
\includegraphics[width=0.35\textwidth]{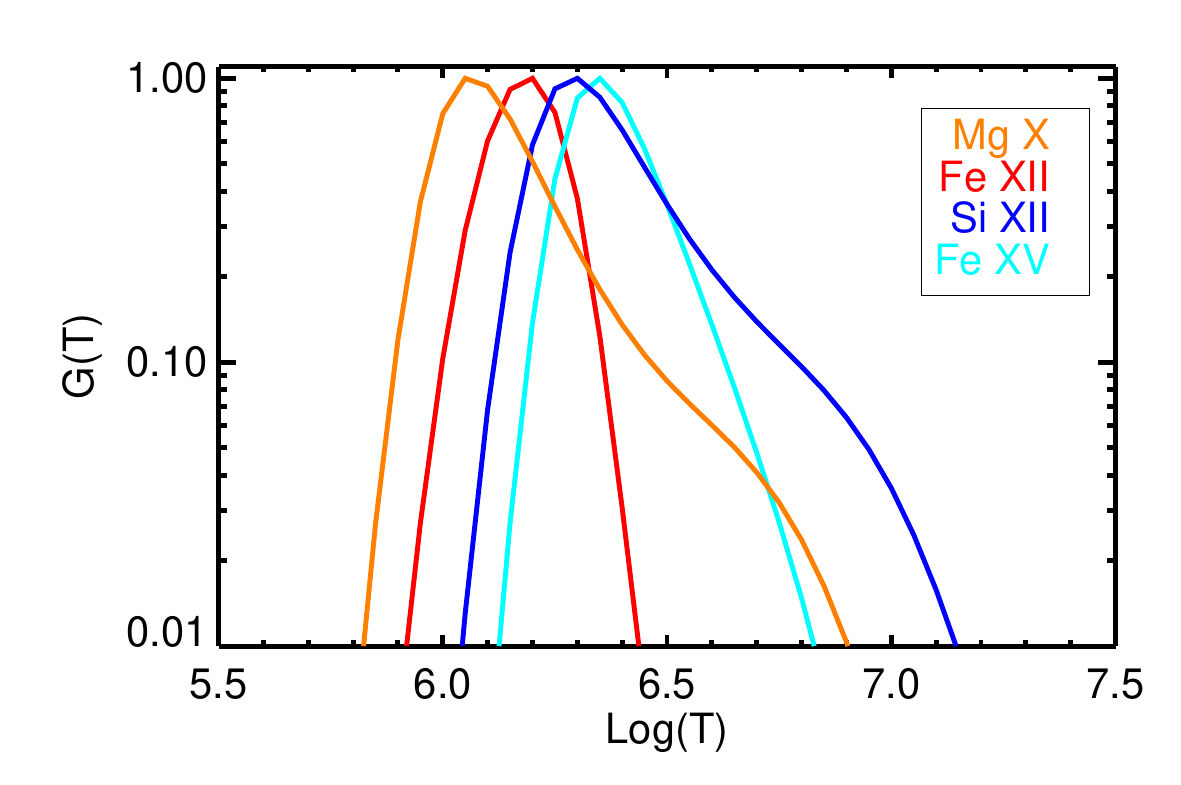}
\caption{\textit{Emissivity functions G(T) normalized to their maxima (from Chianti; Dere \textit{et al.}1997) for relevant EIS lines (Fe~{\sc xii} 195~\AA\  and Fe~{\sc xv} 284~\AA\ and EVE lines (Mg~{\sc x} 624~\AA\ and Si~{\sc xii} 499~\AA; these Li-like ions have response at higher temperatures).
}
        }
\label{fig:flows_gofts}
\end{figure}

We cannot conclusively resolve this tension here but see no reason to doubt the EVE results.
One suggestion is that the EVE hot flows are most visible at higher temperatures (\textit{cf} Figure~\ref{fig:flow_grade_2panel}).
Much of the early EIS work was done with Fe~{\sc xii} 195~\AA, and reported predominantly blueshifts (and also the mixed flow directions seen in the bottom panels of Figure~\ref{fig:bryans}.
\cite{2010A&A...521A..51P} studied Fe~{\sc xv} 284~\AA\ as well, and found examples of regions with very broad wing contributions; this may be a hint that the choice of fit parameters for the high-resolution data (EIS) may obscure the presence of such broad wings.
We note from Figure~\ref{fig:flows_gofts} that Fe~{\sc xii} has a sharply defined  emissivity function G(T), whereas the (Li-like) EVE lines best showing the global effects described here have G(T) values extending to high temperatures, another possible source of bias.

\nocite{1997A&AS..125..149D}

\section{Conclusions}

We have used Sun-as-a-star observations to reveal a general property of active-region coronal plasmas: they contain high-speed prograde flows.
\begin{enumerate}
\item The flows are stable and persistent, varying on time scales of days.
\item The flows are present in coronal lines but not in transition-region or chromospheric lines.
\item There are perceptible differences in the flows at different temperatures, but we cannot yet characterize a detailed temperature dependence.
\item The prograde character of the flows does not depend upon active-region latitude or sunspot cycle.
\end{enumerate}

These results had not been anticipated or predicted in any detail by other EUV spectroscopic observations, which is a puzzle.
The prograde sense of the flows matches that expected for siphon action in the leader-follower magnetic fields of an active region, which may provide a clue towards a physical explanation.
To follow up observationally, we think that image comparisons in detail will be very helpful.
For example, detecting a flow-speed dependence on the tilt angles of active regions would improve our quantitative understanding of the flows and hence help to understand their origin.

It is beyond the scope of this paper to conduct special EIS observations designed to confirm what
EVE sees, but the referee points out that a well-designed differential measurement should be possible,
and also that existing archival EIS data may show the EVE flows with re-analysis.
We note one well-designed data set from the literature: \cite{1983SoPh...87...57T} reported high-resolution observations of the green coronal line of Fe~{\sc xiv}, comparing E and W limbs, and did not see the
effect reported in this paper.
For these observations the minimum height of the (curved) slit was 8700~km above the limb, which may be a clue
to the non-observation in this case.

\section{\bf Acknowledgements:} 
SMM and LF acknowledge support from the UK Research and Innovation's Science and Technology Facilities Council under grant award number ST/T000422/1. 
PCC and TNW are supported by NASA contract number NAS5-02140.
EVE data are courtesy of SDO (NASA) and AIA images as well, via the AIA consortium. 
NOAA Solar Region Summary data were obtained through SolarMonitor.org. 
CHIANTI is a collaborative project involving George Mason 
University, the University of Michigan (USA), University of Cambridge (UK) and NASA 
Goddard Space Flight Center (USA).
The research in this paper originated in the Astro\-nomy Honours Laboratory at the University of Glasgow as a student exercise (co-authors JD, JF, EP, and MS) under the supervision of Prof. Graham Woan.
We thank the referee for helpful comments and also thank Karel Schrijver for reading the draft  manuscript and for finding the Tsubaki reference.

\section{\bf Data Availability Statement:} 

All data used here exist in the public domain: EVE data from the Virtual Solar Observatory  or \url{https://lasp.colorado.edu/eve/data_access/index.html}, and AIA images from the Joint Science Operations Center (JSOC) at \url{http://jsoc.stanford.edu}.
The theoretical database information is from CHIANTI at \url{https://www.chiantidatabase.org}.

\bibliographystyle{mnras}
\bibliography{hotflows}

\end{document}